\documentclass[prd,floats,twocolumn,superscriptaddress]{revtex4-1}
\usepackage{enumerate}
\usepackage{epsf}
\usepackage{graphicx}  
\usepackage{dcolumn}   
\usepackage{bm}        
\usepackage{mathrsfs}
\usepackage{hyperref,graphicx,amsfonts,amssymb,amsthm,amsmath,psfrag}
\newcommand{\be}[1]{\begin{equation} \centering \label{#1}}
\newcommand{\ee}{\end{equation}}
\newcommand{\ba}[1]{\begin{eqnarray} \centering \label{#1}}
\newcommand{\ea}{\end{eqnarray}}

\begin{document}

\title{Higgs-Dilaton cosmology: Universality versus criticality}
\author{Javier Rubio}
\email{javier.rubio@epfl.ch}
\affiliation{Institut de Th\'{e}orie des Ph\'{e}nom\`{e}nes Physiques, 
\'{E}cole Polytechnique F\'{e}d\'{e}rale de Lausanne,  CH-1015 Lausanne, 
Switzerland}
\author{Mikhail Shaposhnikov}
\email{mikhail.shaposhnikov@epfl.ch}
\affiliation{Institut de Th\'{e}orie des Ph\'{e}nom\`{e}nes Physiques, 
\'{E}cole Polytechnique F\'{e}d\'{e}rale de Lausanne, CH-1015 Lausanne, 
Switzerland}

\begin{abstract}
The Higgs-Dilaton model is able to produce an early inflationary expansion followed by a dark energy dominated era responsible for the late time acceleration of the Universe. At tree level, the model predicts 
a small tensor-to-scalar ratio ($0.0021\leq r \leq 0.0034$), a tiny negative running of the spectral tilt ($-0.00057 \leq dn_s/d\ln k \leq -0.00034$) and a nontrivial consistency relation between the spectral tilt of scalar perturbations and the dark energy equation of state, which turns out to be close to a cosmological constant ($0  \leq 1+w_{DE} \leq 0.014$). We reconsider the validity of these predictions in the vicinity of the critical value of the Higgs self-coupling giving rise to
an inflection point in the inflationary potential. The value of the inflationary observables in this case strongly depends on the parameters of the model. The tensor-to-scalar ratio can be large  [$r\sim {\cal O}(0.1)$] and notably exceed 
its tree-level value. If that happens, the running of the scalar tilt becomes positive and rather big [$dn_s/d\ln k \sim {\cal O}(0.01)$] and the equation-of-state parameter of dark 
energy can significantly differ from a cosmological constant [$1+w_{DE}\sim {\cal O}(0.1)$].
\end{abstract}


\maketitle

\section{Introduction}

A B-mode polarization measurement in the cosmic microwave background (CMB) has been recently reported by the BICEP2 collaboration \cite{Ade:2014xna} and 
interpreted as the effect of gravitational waves generated during inflation. Keeping in mind the ongoing debate 
about this interpretation \cite{Flauger:2014qra,Mortonson:2014bja}, if confirmed, it would imply a rather large tensor to scalar ratio $r\sim 0.2$, in conflict with the tree-level predictions 
of one of the simplest and most economical inflationary scenarios: the Higgs inflation model \cite{Bezrukov:2007ep}. 

A way of reconciling the appealing Higgs inflation idea with the BICEP2 result via quantum corrections has been presented 
in Refs.~\cite{Bezrukov:2014bra,Hamada:2014iga} (see also Ref.~\cite{Allison:2013uaa}). Similar approaches involving new physics beyond the standard model can be found in 
 Refs.~\cite{Haba:2014zda,Ko:2014eia,Haba:2014zja,He:2014ora}.
 
In this paper, we reconsider the predictions of a scale-invariant extension of Higgs inflation known as Higgs-Dilaton cosmology \cite{Shaposhnikov:2008xb,Shaposhnikov:2008xi,GarciaBellido:2011de} in light of the recent BICEP2 results. The
paper is organized as follows. In Sec. \ref{sec:model} we review the Higgs-Dilaton model and its tree-level predictions. The effect of quantum corrections 
is discussed in Sec. \ref{sec:QC}, where we introduce a \textit{critical regime} that has not been previously studied. Sec. \ref{sec:CR} is devoted to the detailed analysis of this new regime. The conclusions are presented in 
Sec. \ref{sec:conclusions}.

\section{Tree-level predictions }\label{sec:model}

The Higgs-Dilaton model is a minimalistic scale-invariant extension of Higgs inflation in which the vacuum expectation
 value of the Higgs field is promoted into a singlet scalar field, the dilaton $\chi$, which, together with the Higgs, is nonminimally 
 coupled to gravity.  The late time accelerated expansion of the Universe is implemented within the framework  of unimodular gravity and the standard model 
 is assumed to be a complete theory all the way up to the inflationary scale.
 
 In unimodular gravity \cite{Buchmuller:1988wx,Unruh:1988in} the metric determinant $g$ is restricted to take a constant value, $\vert g \vert=1$. This constraint gives rise 
 to the appearance of an integration constant $\Lambda_0$ at the level of the equations of motion.  The resulting theory is phenomenologically 
 indistinguishable from scale-invariant gravity in the presence of a constant term $\Lambda_0$ \cite{comment1},
\be{general-theory}
\frac{\mathscr L}{\sqrt{-g}}=
\frac{f(\chi,h)}{2}R-
\frac{1}{2}\left(\partial \chi\right)^2-\frac{1}{2}\left(\partial h\right)^2-U(\chi,h)-\Lambda_0 \,.
\ee
This term should be however understood as an initial condition rather than as a fundamental parameter in the action. The prefactor of the Ricci scalar 
and the potential are given respectively by $f(\chi,h)=\xi_\chi \chi^2 +\xi_h h^2$ and 
\begin{eqnarray}
U(\chi,h)&=&\frac{\lambda}{4}
\left(h^2-\frac{\alpha}{\lambda}\chi^2 \right)^2\,, \label{potJ}
\end{eqnarray}
with $h$ denoting the radial component of the Higgs field in the unitary gauge, $\lambda$ its self-coupling 
and $\alpha\sim v^2/M_P^2$ a dimensionless parameter reproducing the hierarchy between the electroweak  $v= 250$ GeV and the Planck scale $M_P=(8\pi G)^{-1/2}=2.44\times 10^{18}$ GeV. 

The phenomenological consequences of the model  can be easily analyzed by performing a conformal transformation $\tilde g_{\mu\nu}=\kappa^{2}f(\chi,h) g_{\mu\nu}$ 
to the so-called Einstein frame, together with a rather involved field redefinition which, in the limit $\xi_\chi\ll \xi_h$, reads  \cite{GarciaBellido:2011de}
\begin{align} \label{rhophi} 
 &\tanh\left[a\kappa(\phi_0-\vert\phi\vert) \right]=\sqrt{1-\varsigma}\cos\theta\,,   \\
 & e^{\gamma \kappa \rho}=\kappa \, \sqrt{ (1+6\xi_\chi)\chi^2+
(1+6\xi_h)h^2}\,.
\end{align}
The constant $\kappa$ in the previous expressions is the inverse of the reduced Planck mass $M_P$ and  we have defined
\begin{align}
&\varsigma\equiv \frac{(1+6\xi_h)\xi_\chi}{(1+6\xi_\chi)\xi_h}\,,\hspace{2mm}     a\equiv \sqrt{\frac{\xi_\chi(1-\varsigma)}{\varsigma}} \,, \hspace{2mm} \gamma\equiv \sqrt{\frac{\xi_\chi}{1+6\xi_\chi}} \,,\nonumber   \\ 
&\tan \theta\equiv \sqrt{\frac{1+6\xi_h}{1+6\xi_\chi}}\frac{h}{\chi} \,, \  \  \ \tan \left[ a\kappa \phi_0\right]\equiv\sqrt{1-\varsigma}\,.\nonumber  
\end{align}
In terms of the new field variables $(\rho,\phi)$, the Lagrangian density takes the very simple form \cite{GarciaBellido:2011de}
\be{angul-theory}
\frac{\tilde{\mathscr L}}{\sqrt{-\tilde g}}=\frac{M_P^2}{2}\tilde R -
\frac{e^{2b\left(\phi\right)}}{2} (\partial \rho)^2- 
\frac{1}{2}(\partial \phi)^2-\tilde V(\rho,\phi)\ ,
\ee
with $e^{2b\left(\phi\right)}\equiv\varsigma\cosh^2\left[a \kappa\left(\phi_0-\vert\phi\vert \right)\right]$.
The potential $\tilde V$ is the sum of two pieces, associated respectively  to the scale-invariant Higgs potential \eqref{potJ} 
\begin{equation}
\tilde U(\phi)=\frac{\lambda M_P^4}{4\xi_h^2 (1-\varsigma)^2}\
\left(1-e^{2b\left(\phi\right)}\right)^2\,, \label{potE}
\end{equation}
and to the unimodular integration constant $\Lambda_0$,
\begin{equation}
\tilde U_{\Lambda}(\rho,\phi)=\frac{\Lambda_0}{\gamma^4} e^{4b\left(\phi\right)} e^{-4\gamma \kappa \rho}\, \label{potEL}\,.
\end{equation}
In the absence of the $\Lambda_0$ term, the field $\rho$ is massless, as corresponds to the Goldstone boson associated to the spontaneous breaking of scale invariance. Motivated by this, 
we will refer to $\rho$ and $\phi$ as the dilaton and the Higgs field respectively.

An important property of $\tilde U(\phi)$ is its flatness at sufficiently large values of  $\phi$. This behavior allows for an inflationary expansion of the Universe in which the scale-breaking potential \eqref{potEL} does not play a significant role  \cite{GarciaBellido:2011de}. Denoting by $N$ the number of e-folds of inflation, the inflationary observables associated to $\tilde U(\phi)$ read \cite{GarciaBellido:2011de}
\begin{eqnarray}
A_s&\simeq&\frac{\lambda \sinh^2\left(4\xi_\chi
N\right)}{1152\pi^2\xi_\chi^2\xi_h^2}
\label{ampcal}\;,\\
n_s&\simeq& 1-8\xi_\chi\coth\left(4\xi_\chi
N\right)\;,\label{tiltcal}\\
\alpha_s&\simeq &-32\xi_\chi^2 \sinh^{-2}\left(4\xi_\chi N\right)\;,\label{runcal}\\
r&\simeq& 192\xi_\chi^2 \sinh^{-2}\left(4\xi_\chi N\right)\label{rcal}\;,
\end{eqnarray}
at leading order in the nonminimal couplings $\xi_\chi$ and $1/\xi_h$. Note that, while the amplitude of the 
scalar perturbations $A_s$ depends on both $\xi_h$ and $\xi_\chi$, the spectral tilt $n_s$, the 
running $\alpha_s\equiv dn_s/d\ln k$ and the tensor-to-scalar ratio $r$ are functions of $\xi_\chi$ only.

At the end of inflation, the Higgs field oscillates around the minimum of \eqref{potE}, releasing its energy into the standard model particles through  
a rather complicated \textit{combined preheating} mechanism  \cite{Bezrukov:2008ut,GarciaBellido:2008ab} in which no dilaton particles are produced \cite{GarciaBellido:2012zu} (see also Refs.  \cite{Enqvist:2014tta,Obata:2014qba}). The 
Higgs field settles eventually down to the minimum of the inflationary potential \eqref{potE} and we 
 are left with a single degree of freedom, the dilaton field $\rho$. At the end of preheating, the potential \eqref{potEL}  takes the form of a dark-energy (DE) quintessence potential
\begin{equation}\label{potDE}
\tilde U_\Lambda(\rho)\simeq \frac{\Lambda_0}{\gamma^4} e^{-4\gamma \kappa \rho}\,.
\end{equation}
Like the inflationary observables \eqref{tiltcal}-\eqref{rcal}, this potential depends only on the nonminimal coupling $\xi_\chi$. This fact allows us to derive a full set of  consistency relations between 
the early and late Universe observables, and in particular between the DE equation-of state parameter associated to \eqref{potDE},
\begin{equation}\label{wxi}
1+w_{ DE}\simeq\frac{8}{3}\frac{\xi_\chi}{1+6\xi_\chi}\,,
\end{equation}
and the spectral tilt of inflationary perturbations \eqref{tiltcal}. This relation reads \cite{GarciaBellido:2011de}
\begin{equation}\label{consistencyC}
n_s\simeq 1-\frac{12(1+w_{ DE})}{5+9w_{ DE}}\coth\frac{6N (1+w_{ DE})}{5+9w_{ DE}}\;.
\end{equation}
The comparison of the tree-level results \eqref{ampcal} and \eqref{tiltcal} with the Planck+WP+highL+BAO $2\sigma$ bounds \cite{Ade:2013zuv},
\begin{align}\label{obsbound}
&\hspace{-5mm} 10^{9}A_s=2.20\pm 0.11\,, \hspace{2mm} n_s= 0.9608\pm 0.0108\,,
\end{align}
puts important constraints on the two nonminimal couplings of the scalar fields to gravity,
\begin{align}\label{chibound} 
&0\lesssim \xi_\chi\lesssim0.0054\,,\hspace{4mm} 48300\lesssim \frac{\xi_h}{\sqrt{\lambda}}\lesssim 68000\,.
\end{align}
These bounds translate, through Eqs. \eqref{runcal}, \eqref{rcal} and \eqref{wxi}, into a tree-level prediction for the tensor-to-scalar ratio, the running of the spectral tilt, and the DE equation-of-state parameter 
\begin{eqnarray}
0.0021\leq &r& \leq 0.0034\,, \label{predr}\\ 
-0.00057 \leq &\alpha_s&\leq -0.00034\,, \label{predrun} \\ 
0  \leq &1+w_{DE}& \leq 0.014\,. \label{predw}
\end{eqnarray}
Note that the value of $w_{DE}$ turns out to be very close to that associated to a cosmological constant ($w_{CC}=-1$) and it is well within the present  Planck+WP+BAO 2$\sigma$ observational bound \cite{Ade:2013zuv},
\begin{equation}
\label{obsw}
-0.38 \leq 1+w_{obs} \leq 0.11\,.
\end{equation}

\section{Quantum corrections}\label{sec:QC} 

The predictions and consistency relations derived in the previous section are subject to changes in the presence of quantum corrections. 
A regularization procedure respecting the symmetries of the classical action \eqref{general-theory} together with the approximate shift symmetry of the dilaton field in the Einstein frame 
was presented in Ref. \cite{Bezrukov:2012hx}  (see also Refs.  \cite{Bezrukov:2008ej,Bezrukov:2009db,Bezrukov:2010jz}). In that prescription, the constant 't Hooft-Veltman 
parameter of dimensional regularization is replaced by a combination of the 
scalar fields $h$ and $\chi$ with the appropriate dimension, 
$\mu^2\rightarrow \mu^2(\chi,h)$. The quantity $\mu^2(\chi,h)$ is defined in the original frame \eqref{general-theory} and its precise form cannot be completely determined in the absence of an 
ultraviolet completion of gravity. The quantization of the Higgs-Dilaton model requires the choice of a classical action together with a choice of 
subtraction rules, which should be understood as the remnant of a given ultraviolet completion. The 
simplest possibility (albeit nonunique, cf. Refs. \cite{Barvinsky:2008ia,DeSimone:2008ei,Barvinsky:2009ii}) is to take $\mu^2\propto \xi_\chi\chi^2+\xi_h h^2$ or equivalently $\tilde \mu\propto M_P^2$ in the Einstein frame \cite{Bezrukov:2012hx}. 
\begin{figure}
\centering
\includegraphics[scale=0.34]{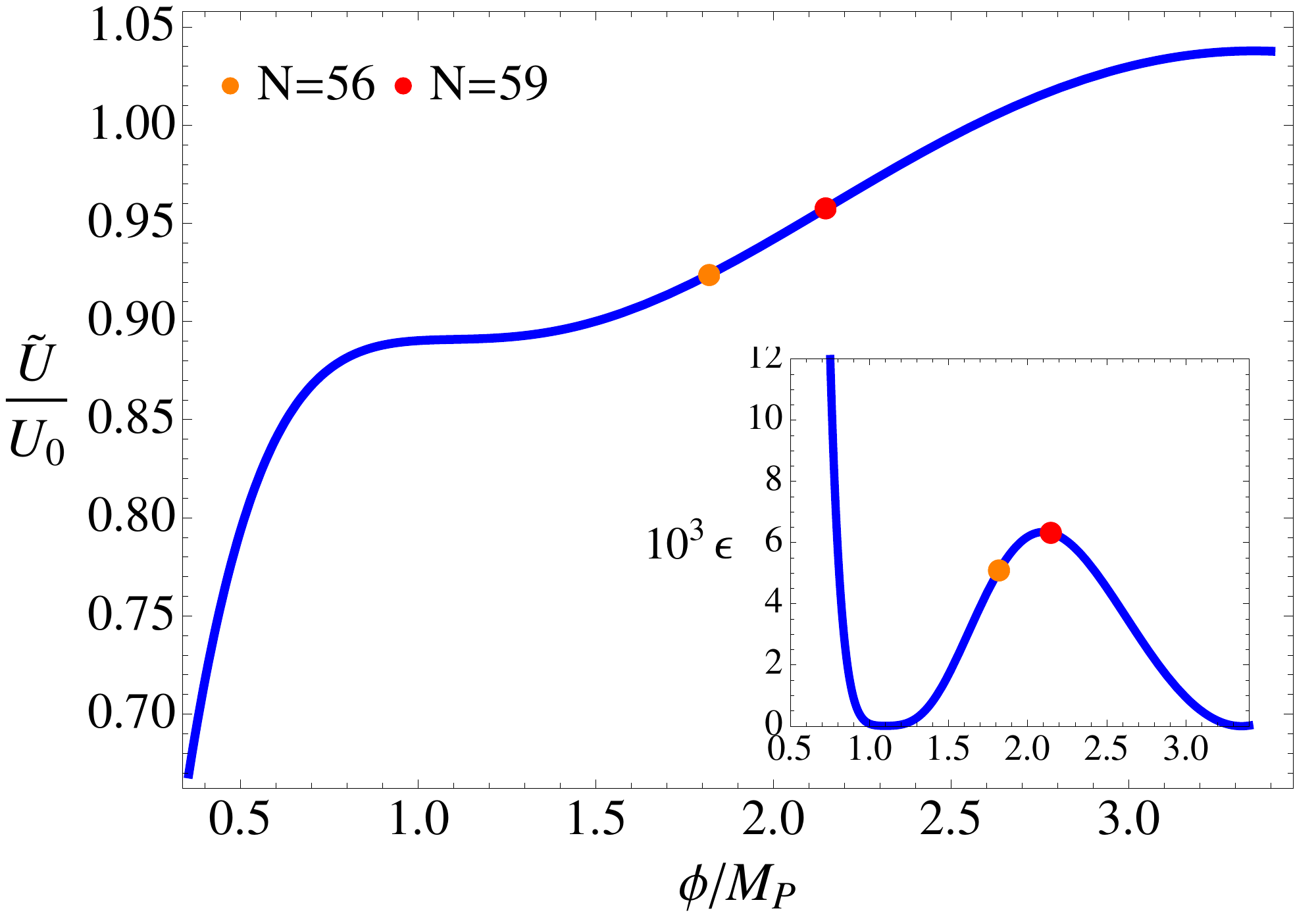}
\caption{(color online) The shape of the renormalization group enhanced potential in the critical regime in which $\lambda_0-b/16\simeq 0$. The subfigure shows the nonmonotonic 
behavior  of the slow-roll parameter $\epsilon$ for such a potential.} \label{NMpot}
\end{figure}

The renormalization group enhanced potential in the Einstein frame reads
 \be{defin-potQ}
\tilde U(\phi)=\frac{\lambda (\tilde \mu) M_P^4}{4\xi_h^2 (1-\varsigma)^2}
\left(1-e^{2b\left(\phi\right)}\right)^2\,,
\ee
with the effective self-coupling $\lambda(\tilde \mu)$ given by
\begin{equation}
\label{Lrun}
\lambda(\tilde \mu) =\lambda_0+b \log^2 \left(\frac{\kappa\, \tilde \mu}{q_{\rm eff}}\right)\,.
\end{equation}
Here $b\simeq 2.3 \times 10^{-5}$ and  $\lambda_0$ is  some function of the top quark pole mass, the Higgs mass and the strong coupling constant \textit{at the inflationary scale} \cite{comment2}, whose precise form 
will not be relevant for the present discussion \cite{Bezrukov:2014bra}. The scale 
\begin{equation}\label{muscale}
 \tilde \mu^2=\frac{M_P^2\left(1-e^{2b\left(\phi\right)}\right)}{(1-\varsigma)}
\end{equation}
is proportional to the top quark mass in the Einstein frame. The proportionality constants, together with some functional dependence on 
the top quark pole mass, the Higgs mass and the strong coupling constant \textit{at the inflationary scale}  \cite{Bezrukov:2014bra}, have been incorporated in the definition of $q_{\rm eff}$.

The inspection of the renormalization group enhanced potential \eqref{defin-potQ} reveals the existence of three different regimes:
\begin{enumerate}[i)] 
\item {\it Universality regime}.---For $\lambda_0 \gg b/16$, the form of the potential is almost independent of 
the precise values of the parameters $\xi_\chi$, $\xi_h$ and $q_{\rm eff}$ appearing within the logarithmic correction in \eqref{Lrun}. As in the tree-level case, 
the potential effectively depends on two parameters ($\sqrt{\lambda}/\xi_h$ and $\xi_\chi$), which can be fixed with the two observational bounds in \eqref{obsbound}.
\item{\it Critical regime}.--- If $\lambda_0=b/16$, both the first and the second derivative of the potential are equal to zero at some field value along the inflationary slow-roll evolution.
\item {\it Forbidden regime}.---If $\lambda_0 \lesssim b/16$, the potential develops a wiggle and inflation is no longer possible. Smaller values of
$\lambda_0$ make the electroweak vacuum unstable \cite{Bezrukov:2012sa,Degrassi:2012ry,Buttazzo:2013uya}.
\end{enumerate}
\section{The  critical regime}\label{sec:CR}
 A detailed analysis of the {\it universality regime} in the Higgs-Dilaton model was performed in Ref.~\cite{Bezrukov:2012hx}.  As expected, the values of the inflationary observables were shown to be 
 stable and to coincide with the tree-level estimates \eqref{ampcal}-\eqref{rcal}. The consistency relation \eqref{consistencyC} and the 
 bounds \eqref{predr}-\eqref{predw}  became in this case solid predictions, valid even in the presence of quantum corrections.
 \begin{figure}
\includegraphics[scale=0.3]{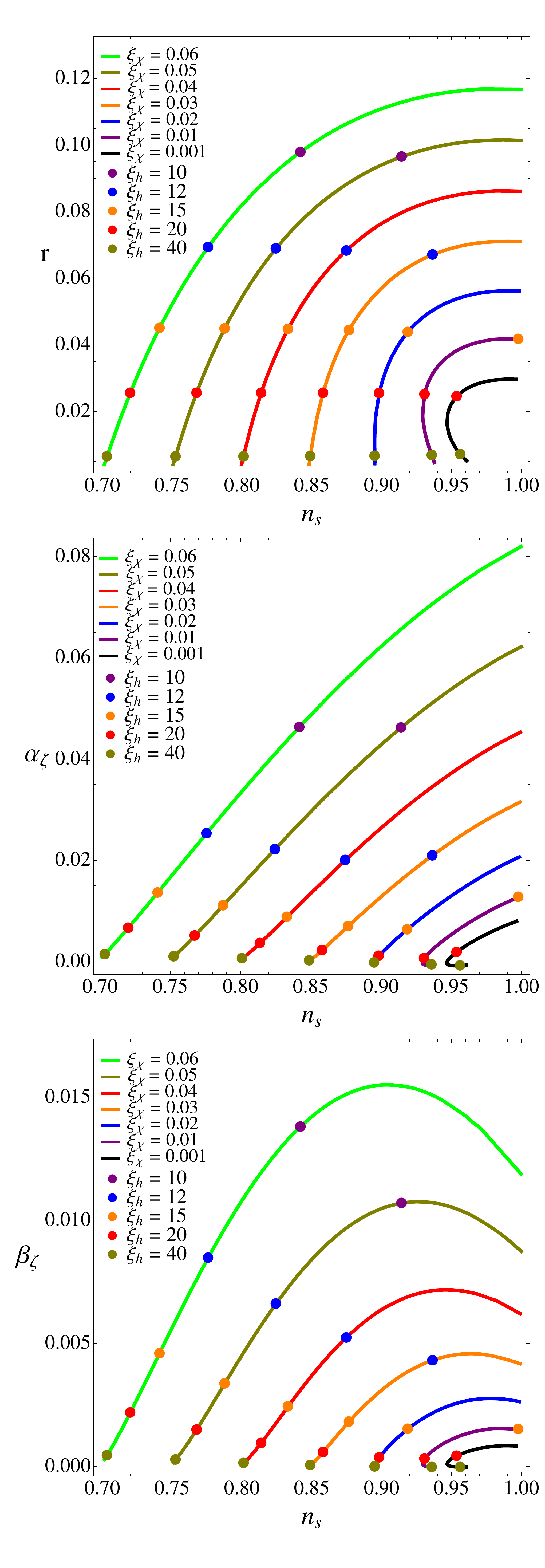}
\caption{(color online)  Dependence of the inflationary observables on the nonminimal couplings $\xi_h$ and $\xi_\chi$ for $q_{\rm eff}=1.05$. Different lines are associated 
to particular values of $\xi_\chi$. The value of $\xi_h$ along these lines varies within the interval $\lbrace9,50\rbrace$.}
 \label{combined}
\end{figure}

The situation in the vicinity of the critical point $\lambda_0=b/16$ is rather different.  The dependence of the running function $\lambda(\tilde \mu)$ on the parameters $\lambda_0$, $\xi_\chi$, $\xi_h$ 
and $q_{\rm eff}$ is now essential and the shape of the inflationary potential strongly differs from the tree-level case (cf. Fig.~\ref{NMpot}). The slow-roll parameter $\epsilon$ becomes nonmonotonic, opening 
the possibility of getting a sizable tensor-to-scalar ratio \cite{BenDayan:2009kv,Hotchkiss:2011gz}. To illustrate this point, we will use the COBE normalization to fix the value of $\lambda_0$ 
and will reduce the space of parameters by setting $q_{\rm eff}=1.05$ \cite{comment3}. 
 The nonminimal coupling $\xi_h$ will be taken to be a free parameter and 
the range of variation of $\xi_\chi$ will be restricted to the range dictated by  the low-energy observational bound \eqref{obsw}, the theoretical requirement $\xi_\chi>0$ and the relation \eqref{wxi}, namely $0\leq \xi_\chi\lesssim 0.06$. 

Numerical results for the tensor-to-scalar ratio $r$, the running of the spectral tilt $\alpha_s\equiv d \ln n_s/d\ln k$ and the running of the running $\beta_s\equiv d^2\ln n_s/d \ln k^2$ as a function of the spectral tilt $n_s$ are displayed in Fig.~\ref{combined} 
 for varying values of the nonminimal couplings $\xi_\chi$ and $\xi_h$.  As shown in the upper panel, it is possible to obtain any value of $r$ and $n_s$ by properly 
 choosing  the values of   $q_{\rm eff}$ and the nonminimal couplings $\xi_h$ and $\xi_\chi$.  For the sizable values of $r$ suggested by the BICEP2 collaboration \cite{Ade:2014xna},
 the runnings $\alpha_s$ and $\beta_s$ (2nd and 3rd panel respectively) turn out to be positive and rather large \cite{comment4}. 
  The tree-level consistency condition \eqref{consistencyC} does no longer apply and  the equation-of-state parameter of dark energy [cf. Eq. \eqref{wxi}] can significantly differ from a cosmological constant [$1+w_{DE}\sim {\cal O}(0.1)$] without spoiling inflation \cite{comment7}.

The value of the nonminimal coupling $\xi_h$ in the critical case can be rather small, $\xi_h\sim{\cal O}(10)$, which has interesting consequences for the range of validity of the theory. As shown in Ref. \cite{Bezrukov:2012hx},  perturbation theory,  when performed around the constant electroweak vacuum $v$, breaks down at a scale $\Lambda_G \sim M_P/\xi$, where an ultraviolet completion of the model or a self-healing mechanism should be expected. The small value of $\xi_h$ in the critical region delays the onset of this strong coupling regime and avoids the introduction of an ultraviolet completion at energy scales significantly below the Planck scale.  
 
\begin{figure}
\centering
\includegraphics[scale=0.36]{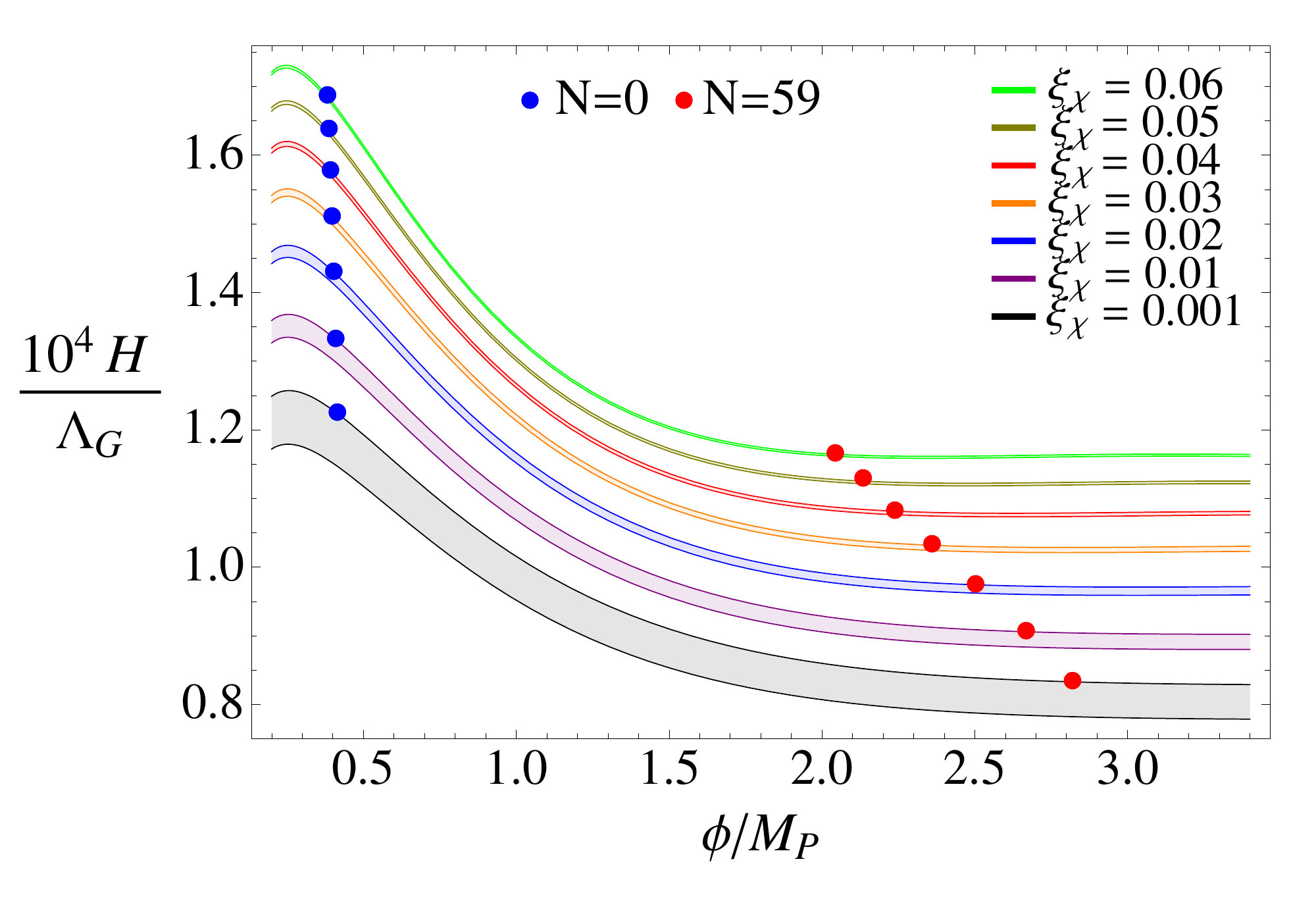}
 \vspace{-3mm}
\caption{(color online) Ratio between the typical momenta $k \sim H$ of the scalar perturbations produced during inflation and the gauge cutoff $\Lambda_G$
 as a function of the Higgs field $\phi$. Different colors correspond to different values of $\xi_\chi$. Shaded areas account for changes of $\xi_h$ leading
  to a spectral tilt in the range $n_s \in \lbrace 0.95,0.97 \rbrace $.}
 \label{cutoffplot}
\end{figure}

  The self-consistency of the model is also guaranteed when perturbations are computed around inflationary Higgs field values \cite{comment5}. As shown in Fig.~\ref{cutoffplot}, the typical energy of the scalar perturbations  produced during critical inflation, $H \sim {\cal O}(\sqrt{\lambda} M_P/\xi_h)$ is well below the cutoff of the theory at those energies, $\Lambda_G\sim {\cal O}(M_P/\sqrt{\xi_h})$ \cite{comment6}.
\vspace{-3mm}
\section{Conclusions}\label{sec:conclusions}

We have clarified the effect of quantum corrections on the predictions of the Higgs-Dilaton model. Away 
from the critical point, the inflationary and dark energy observables are effectively controlled by the nonminimal 
coupling of the dilaton field to gravity. The consistency of the model requires a small tensor to scalar ratio, a tiny negative 
running of the spectral tilt and a dark energy equation of state close to a cosmological constant. 
 
In the vicinity of the critical point, the inflationary observables strongly depend on the nonminimal couplings of the Higgs and 
the dilaton field to gravity and on the \textit{inflationary} Higgs and top quark masses encoded in the value of $q_{\rm eff}$. The model does not require the inclusion of any cutoff scale below the Planck scale and can give rise to sizable values of 
the tensor-to-scalar ratio. For those values, the running of the scalar tilt (and the running of the running) becomes positive and rather large.  Suitable choices of the nonminimal coupling $\xi_\chi$ allow for a late time evolution of the Universe that can significantly differ from a de Sitter expansion. A proper comparison of our results with observations would require a complete fit of the inflationary spectra to the CMB data along the lines presented in Refs. \cite{Lesgourgues:2007gp,Lesgourgues:2011re,Blas:2011rf}. That analysis is beyond the scope of this paper.
\section*{ACKNOWLEDGMENT}

This work was partially supported by the Swiss National Science
Foundation.


\begin{thebibliography}{99}
\bibitem{comment1} 
We omit the SM Lagrangian not involving the Higgs potential.
\bibitem{comment7}
 Note that, contrary to the consistency condition \eqref{consistencyC}, Eq. \eqref{wxi} is a low-energy relation that holds even in the presence of quantum corrections.
\end{thebibliography}
\end{document}